\documentclass[aps,longbibliography,twocolumn,letterpaper,10pt,showkeys,superscriptaddress,floatfix]{revtex4-1}

\usepackage{amsmath,amssymb}
\usepackage{textcomp,gensymb}
\usepackage{graphicx}
\usepackage{dcolumn}
\usepackage{bm}
\usepackage{soul}
\usepackage{comment}

\usepackage{float}
\usepackage{color}
\usepackage{hyperref}

\usepackage[table]{xcolor}
\usepackage{siunitx}

\usepackage[normalem]{ulem}

\newcommand{\jp}[1]{{\color{black}#1}}

\usepackage{lipsum}

\setstcolor{red}
\begin{document}

\preprint{APS/123-QED}

\title{Vibrational similarities in jamming-unjamming of polycrystalline and disordered granular packings}

\author{Juan C. Petit}
\email{corresponding author: Juan.Petit@dlr.de}
\author{Saswati Ganguly}
\affiliation{Institute of Materials Physics in Space, German Aerospace Center (DLR), 51170 K\"oln, Germany}
\author{Matthias Sperl}
\affiliation{Institute of Materials Physics in Space, German Aerospace Center (DLR), 51170 K\"oln, Germany}
\affiliation{Institut f\"{u}r Theoretische Physik, Universit\"{a}t zu K\"{o}ln, 50937 K\"{o}ln, Germany}%

\date{\today}

\begin{abstract}

We investigate the vibrational properties of polycrystalline monodisperse and disordered bidisperse granular packings during jamming and unjamming using discrete element method simulations. Both systems deviate from Debye scaling at low frequencies $(\omega)$, but only bidisperse packings exhibit a low-$\omega$ plateau. The low $\omega$ exponent ($\alpha$) in bidisperse packings evolves smoothly from zero (plateau) to near one (Debye scaling) with increasing packing fraction, whereas in polycrystalline packings, it changes discontinuously near jamming/unjamming, due to the nature of the contact network rearrangements. Despite structural modifications during the compression-decompression cycle, the exponent remains unchanged at the same distance from jamming density, regardless of the history. Nonaffine displacements and contact orientational order further confirm that structural features that impact low-$\omega$ vibrational states and, hence, mechanical properties are largely restored upon decompression, reinforcing vibrational similarities between jamming and unjamming states.

\end{abstract}

\maketitle


The phenomenon of jamming in granular media is critical for understanding the transition between fluid-like and solid-like states in granular packings \cite{majmudar2007jamming,ohern2003jamming,donev2004jamming,petit2020additional,kumar2016memory,vagberg2010glassines,Hecke2009Jamming,Silbert2005Vibrations,Goodrich2014Solids,Zhang2021Disorder,Tong2015From}. This transition is not merely a theoretical concern; it has significant implications for practical applications across various fields, including material science, geophysics, industrial processes, and cell biology \cite{amend2016soft, zuriguel2005jamming,jiang2014robotic, peters2015dynamic,kostynick2022rheology, lawson2021jamming, atia2021cell, oswald2017jamming}. As granular systems undergo jamming, they acquire rigidity, a property traditionally associated with crystalline solids, which exhibit long-range order \cite{Forster,Martin_Parodi_Pershan}. However, disordered solids, which lack this long-range order, also resist external stresses, challenging our conventional understanding of rigidity and mechanical stability~\cite{sethna}.

Previous theoretical and experimental studies have explored the mechanical transition of monodisperse crystals to disordered solids \cite{Tong2015From, jin2010first, hanifpour2015structural, Goodrich2014Solids}. In three-dimensional systems, a key feature of this transition is the emergence of a coexistence region in the mechanical coordination number (Z), appearing as a plateau from $\phi_J \approx 0.64$  to  $\phi \approx 0.68$  at  $Z \approx 6$, marking the isostaticity of the disordered structure before it transitioned to an ordered state \cite{jin2010first}. Supporting this, studies have identified intermediate polycrystalline states with disordered-like mechanical properties near jamming, which evolve into crystalline behaviour under high pressure \cite{Goodrich2014Solids}. This coexistence region between ordered and disordered packings has been confirmed experimentally and validated through simulations, providing strong evidence of a structural transition \cite{hanifpour2015structural}. Furthermore, simulations in both two and three dimensions reveal an order-disorder transition displaying an intermediate polycrystalline phase with distinct mechanical properties \cite{Tong2015From}.
Beyond mechanical behaviour, the vibrational density of states (DOS) also reflects these structural changes. In a 2D triangular crystal at a fixed packing fraction $\phi$, the DOS is highly sensitive to increasing disorder. As disorder grows, it transitions from a crystalline profile (low disorder) to a polycrystalline form (intermediate disorder) and eventually to a fully disordered structure (high disorder). These findings demonstrate that packing structure is directly mirrored in the DOS, particularly at low frequencies, where soft modes, including the boson peak, become dominant.

Although the effect of disorder on the mechanical properties of both polycrystalline and disordered packings has been explored \cite{donev2004jamming, Goodrich2014Solids, Zhang2021Disorder, Tong2015From}, The change in vibrational behaviour during jamming, where particle contacts quickly increase at a certain density, and during unjamming, where particle contacts suddenly decrease at a similar density, has not been fully studied, especially when comparing these two states in a compression-decompression cycle. A compression-decompression protocol can significantly alter the packing structure, thereby impacting the mechanical properties of the system \cite{kumar2016memory, hanifpour2015structural, klumov2014structural}. These structural changes are expected to be directly reflected in the DOS, capturing the influence of packing modifications on vibrational modes and mechanical behaviour. Examination of the difference in vibrational behaviour of the packing near jamming and unjamming states, with similar packing fractions but different structural arrangements, can offer important conceptual insights.
Most research focuses on either jamming or unjamming, often overlooking the interplay between the two. Comparing polycrystalline monodisperse and disordered bidisperse structures helps clarify whether the soft vibrational modes arise solely from disorder or if they also emerge in systems with partial order (polycrystals). This distinction offers deeper insights into the structural features that govern mechanical behaviour.

Our objective is to investigate how the vibrational behaviour of two-dimensional (2D) polycrystalline monodisperse and disordered bidisperse packings evolve along the jamming and unjamming path,
with a focus on how structural changes affect the corresponding DOS. 2D systems are easier to replicate in experimental setups, including microgravity, facilitating various empirical and fundamental studies that validate our theoretical perspective \cite{fischer2021force, grasselli2017dynamics}. Insights from simulations of compression and decompression during the jamming-unjamming transition in 2D granular solids can deepen our understanding of seismic wave propagation \cite{fischer2021force}, acoustic wave behavior \cite{tell2020elastic, tell2020acoustic}, soil compaction, and granular flow in industrial applications \cite{muller2023space, d2024rheological}.

\begin{figure}[t]
    \centering \includegraphics[scale=0.42]{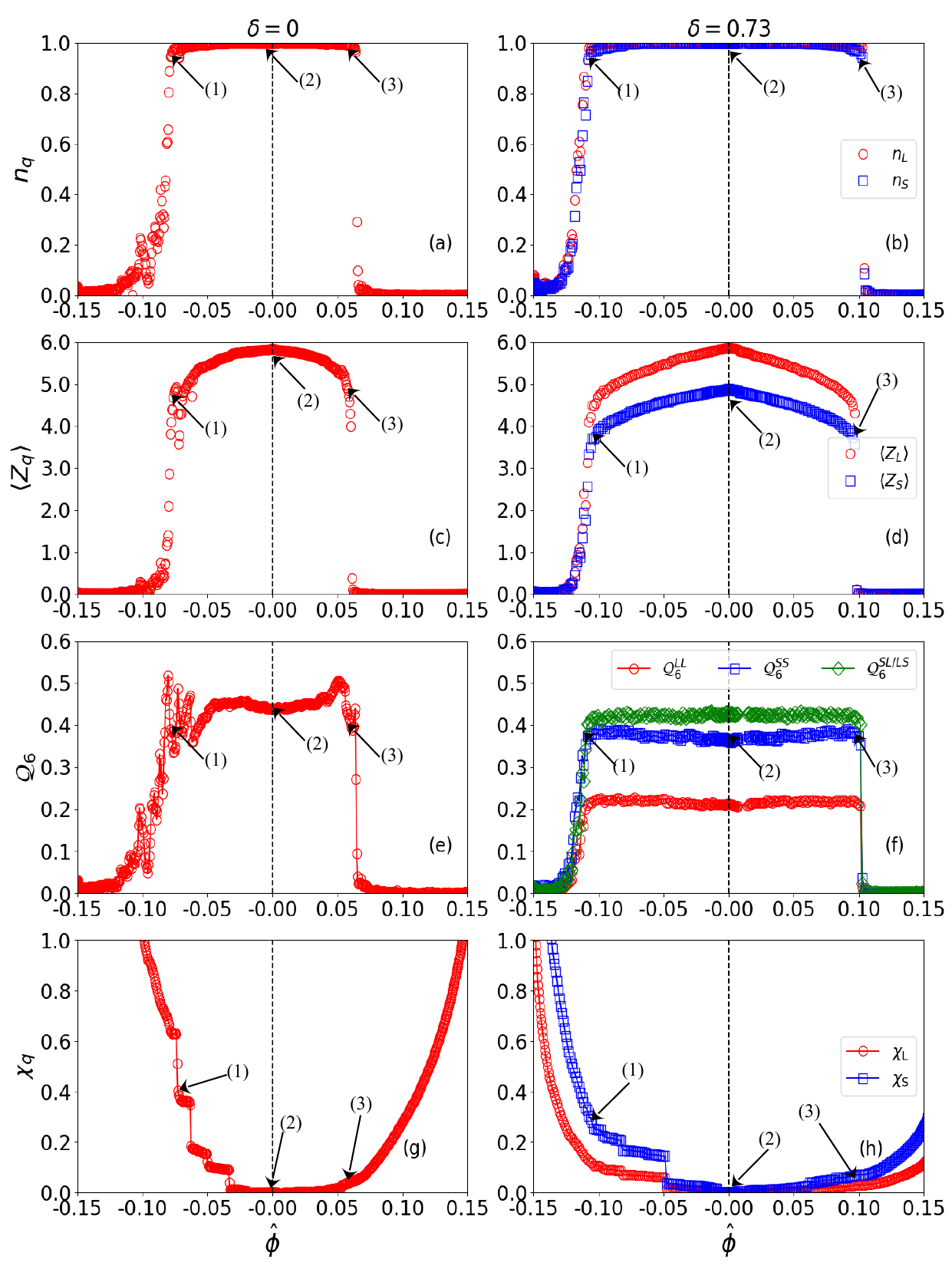}
    \caption{(a)-(b) Fraction of particles in contact, $n_{q}$, 
    (c)-(d) mean contact number, $\langle Z_{q} \rangle$,
    (e)-(f) contact orientational order, $\mathcal{Q}_{6}$, and
    (g)-(h) the cumulative nonaffine parameter, $\mathcal{X}_{q}$, 
    as a function of $\hat{\phi}$ for monodisperse (left panel) 
    and bidisperse packings (right panel) as the systems are compressed to $\phi_{\mathrm{max}}$ and decompressed. For the bidisperse packing, the $\delta = 0.73$ and $X_{\mathrm S} = 0.5$. 
    Here, $q \in [\text{L}, \text{S}]$, except in the monodisperse case where $n_{\mathrm{L}} = n_{\mathrm{S}}$ by definition.
    $\mathcal{X}_{q}$ is intentionally scaled by 1000 to enhance the readability of the y-axis.
    The arrows indicate (1) near jamming, (2) $\phi_{\rm max}$, and (3) near unjamming. 
    Particle configurations on these states are shown in Fig.~\ref{fig2}. 
    The dashed line corresponds to $\hat{\phi} = \phi -\phi_{\rm max}= 0$ or $\phi = \phi_{\rm max}$. }
	\label{fig0}
\end{figure}

Simulations using the 2D discrete element method are performed with the software MercuryDPM \cite{cundall1979discrete, weinhart2020fast, petit2022bulk}. Monodisperse and bidisperse packings 
are formed by $N = 3000$ particles. The bidisperse packing has a number of large, $N_{\mathrm L}$, 
and small, $N_{\mathrm S}$, particles with radius $r_{\mathrm L}$ and $r_{\mathrm S}$. The size ratio is $\delta = r_{\mathrm{S}}/r_{\mathrm{L}} = 0.73$ with a concentration of small particles of 
$X_{\mathrm S} = N_{\mathrm S} \delta^{2} / (N_{\mathrm L} + N_{\mathrm S} \delta^{2}) = 0.5$. Therefore, $\delta = 0$ represents the monodisperse packing. $\delta = 0.73$ and $X_{\mathrm{S}} = 0.5$ are chosen to prevent long-range order \cite{majmudar2007jamming, ohern2003jamming, morse2016geometric}. Simulations are performed using an isotropic compression–decompression protocol with a linear, frictionless contact model. Periodic Boundary motion is applied smoothly and symmetrically in all directions using a cosine-based strain profile, ensuring continuous deformation and minimal inertial effects (see Sec.~I of the Supplemental Material \cite{SupplementalMaterial} for details). The system is first compressed from an initial packing fraction $\phi_0 = 0.80$ up to a maximum value $\phi_{\mathrm{max}} = 0.95$, then decompressed back to $\phi_0$ at the same rate. The mean volumetric strain rate during the compression–decompression protocol, $\langle\dot{\varepsilon}_{\rm v}\rangle \sim 10^{-4}$, is approximately 100 times higher than in previous work \cite{kumar2016memory}. Despite this, the protocol consistently yields polycrystalline monodisperse and disordered bidisperse packings. Mechanical equilibrium is maintained throughout, with a stable kinetic-to-potential energy ratio of $\sim 10^{-4}$ near the jamming transitions. This process allows for the extraction and analysis of structural variables to assess how monodisperse and bidisperse packings respond to compression and decompression, and their impact on vibrational behaviour during the jamming-unjamming transition.

\begin{figure}[t]
    \centering \includegraphics[scale=0.5]{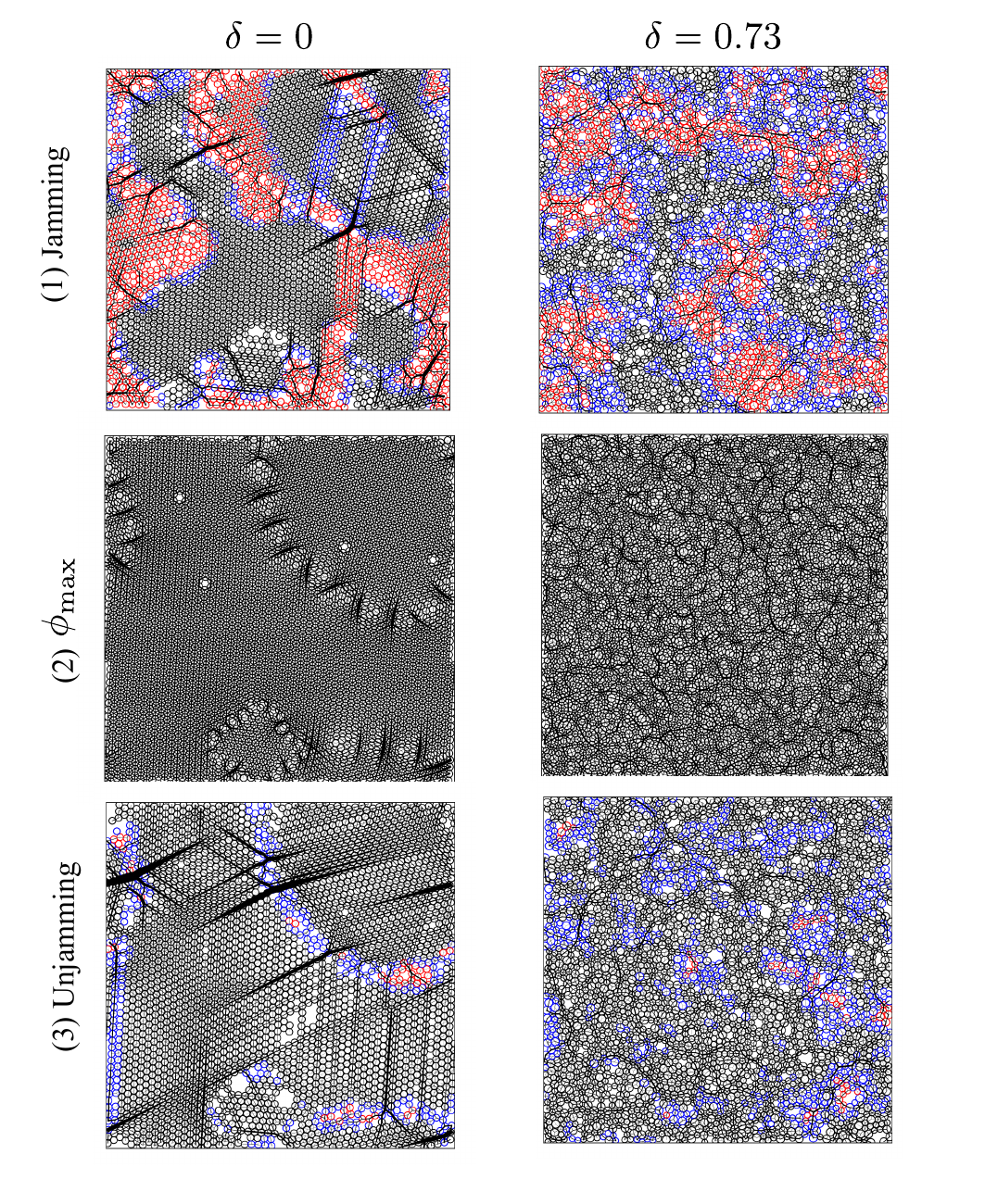}
    \caption{Particle configurations and force chains for monodisperse 
    $(\delta = 0)$ and bidisperse $(\delta = 0.73 ,  X_{\mathrm{S}} = 0.5 )$ 
    packings are shown near jamming (1), at $\phi_{\rm max}$ (2), and near 
    unjamming (3), consistent with Fig.~\ref{fig0}. The width of the force 
    chain lines corresponds to normal forces greater than  $2\langle F_{n} \rangle$, 
    ensuring the visibility of the particle colors. Black, blue, and red circles 
    represent the nonaffine motion of the  $i$-th particle, corresponding to low $( \chi_{i} < 1 )$, intermediate $( 1 \leq \chi_{i} < 5 )$, and high $(\chi_{i} \geq 5)$ values, respectively.}
	\label{fig2}
\end{figure}

Fig.~\ref{fig0} (a)-(b) and (c)-(d) illustrate the fraction of particles in contact, 
$n_{q} = N^{c}_{q}/N_{q}$, and the mean contact number, 
$\langle Z_{q} \rangle = \sum_{i=1}^{N_{q}} Z^{q}_{i} / N_{q}$, as a function of relative packing fraction, $\hat{\phi}=\phi-\phi_{\rm max}$. Here, $q \in [\text{L}, \text{S}]$,  $Z^{q}_{i}$ is the number of contacts for the $i$-th large (L) or small (S) particle, and  $N^{c}_{q}$ is the total contact number for large or small particles. $\hat{\phi}$ distinguishes between compression and decompression branches, with compression defined as $\hat{\phi} = \phi - \phi_{\rm max} \leq 0$ and decompression as $\hat{\phi} = \phi_{\rm max} - \phi > 0$. For monodisperse and bidisperse packings, $n_{q}$ and $\langle Z_{q} \rangle$ exhibit a sharp increase upon compression, with $n_{q} \approx 1$ and $\langle Z_{q} \rangle \sim$ $4$-$5$, at a specific $\hat{\phi}$ consistent with the jamming density. As $\hat{\phi} \to 0$, $n_{q}$ and $Z_{q}$ approaches their respective maximum values of $ 1$ and $6$. During decompression, $n_{q}$ and $\langle Z_{q} \rangle$  drop discontinuously at a different $\hat{\phi}$ than near jamming, indicating the transition to an unjammed structure.

The jamming ($\phi_{J}$) and unjamming ($\phi_{uJ}$) densities are
determined using the five-point finite difference method 
$(\sim O(\Delta \phi^{4}))$ to compute the derivative of the particle fraction, $n_{q}$, 
with respect to  $\phi$. The method provides a pronounced peak at $\phi_{J}/\phi_{uJ}$, 
indicating a rapid shift in particle arrangement that marks the onset of either a jammed or unjammed structure, see Sec.~II in \cite{SupplementalMaterial}. The extracted jamming and unjamming densities for monodisperse packing are $\phi_J \approx 0.87$ and $\phi_{uJ} \approx 0.88$. These values are consistent with the formation of some crystallization inside the packing but far from the triangular lattice structure of $\sim 0.90$ observed in Ref.~\cite{donev2004jamming}. Polycrystals are already formed near the jamming density (1), remain intact at $\phi_{\max}$ (2) after particle rearrangement during further compression, and persist even near the unjamming state (3). Fig.~\ref{fig2} reveals polycrystalline structures with inhomogeneous force chains, where regions of high non-affine particle motion (colored red, $\chi_i \geq 5$) are primarily localized near grain boundaries, delineating distinct local structural domains and highlighting deviations from affine deformation. For the bidisperse packing, the unjamming density is $\phi_{uJ} \approx 0.84$, which is $0.5\%$ higher than $\phi_{J}$. This value agrees with that reported for amorphous bidisperse mixtures in
Refs.~\cite{donev2004jamming,vagberg2010glassines}. The structure near jamming in bidisperse packings appears disordered, as indicated by the more uniform distribution of force chains. This disordered character becomes more evident at the maximum packing fraction, $\phi_{\max}$, where regions of distinct degrees of non-affine motion are more broadly distributed throughout the system, see the right panel of Fig.~\ref{fig2}. Such a structure evolves with smooth incremental changes along compression-decompression paths. Subtle structural changes in both packings during this process are also supported by the structure factors presented in Sec.~III of the Supplemental Material \cite{SupplementalMaterial}.

\begin{figure}[t]
    \centering \includegraphics[scale=0.28]{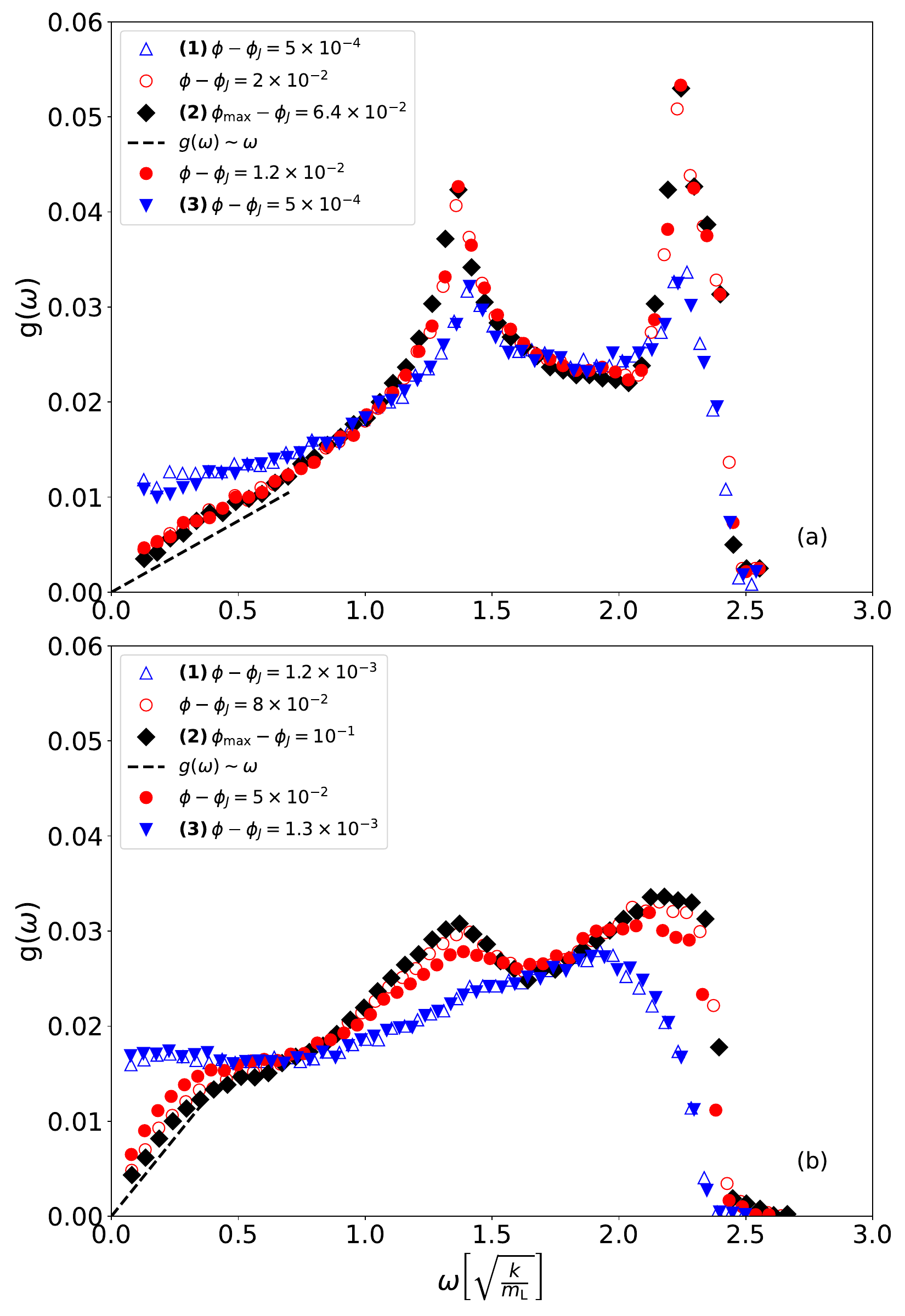}
    \caption{ The DOS, $g(\omega)$ versus $\omega$, for (a) monodisperse and (b) bidisperse packings along the compression-decompression process. Open red and blue ($\triangle$, $\bigcirc$) symbols represent the system during the compression, whereas solid red and blue ($\triangledown$, $\bigcirc$) the decompression at different distances from jamming/unjamming. $\phi_{\rm{max}} = 0.95$ is in black solid ($\Diamond$). The Debye scaling law, $g(\omega) \sim \omega$, for 2D solids is represented in black dashed lines. The jamming and unjamming densities for monodisperse packing are $\phi_J \approx 0.87$ and $\phi_{uJ} \approx 0.88$, while for bidisperse packing, $\phi_J \approx 0.83$ and $\phi_{uJ} \approx 0.84$, respectively.} 
	\label{fig3}
\end{figure}

\begin{figure}[t]
    \centering \includegraphics[scale=0.28]{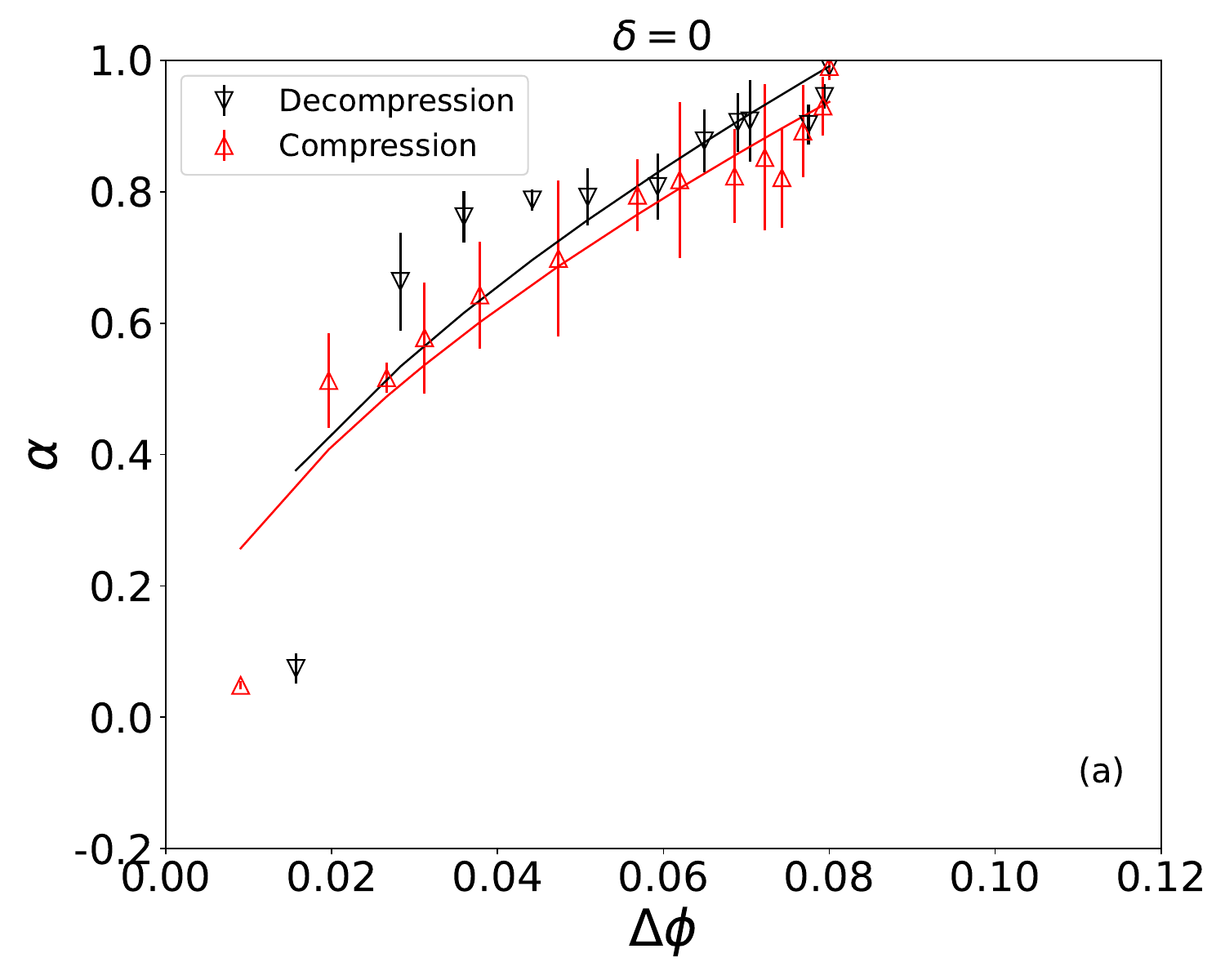}
    \centering \includegraphics[scale=0.28]{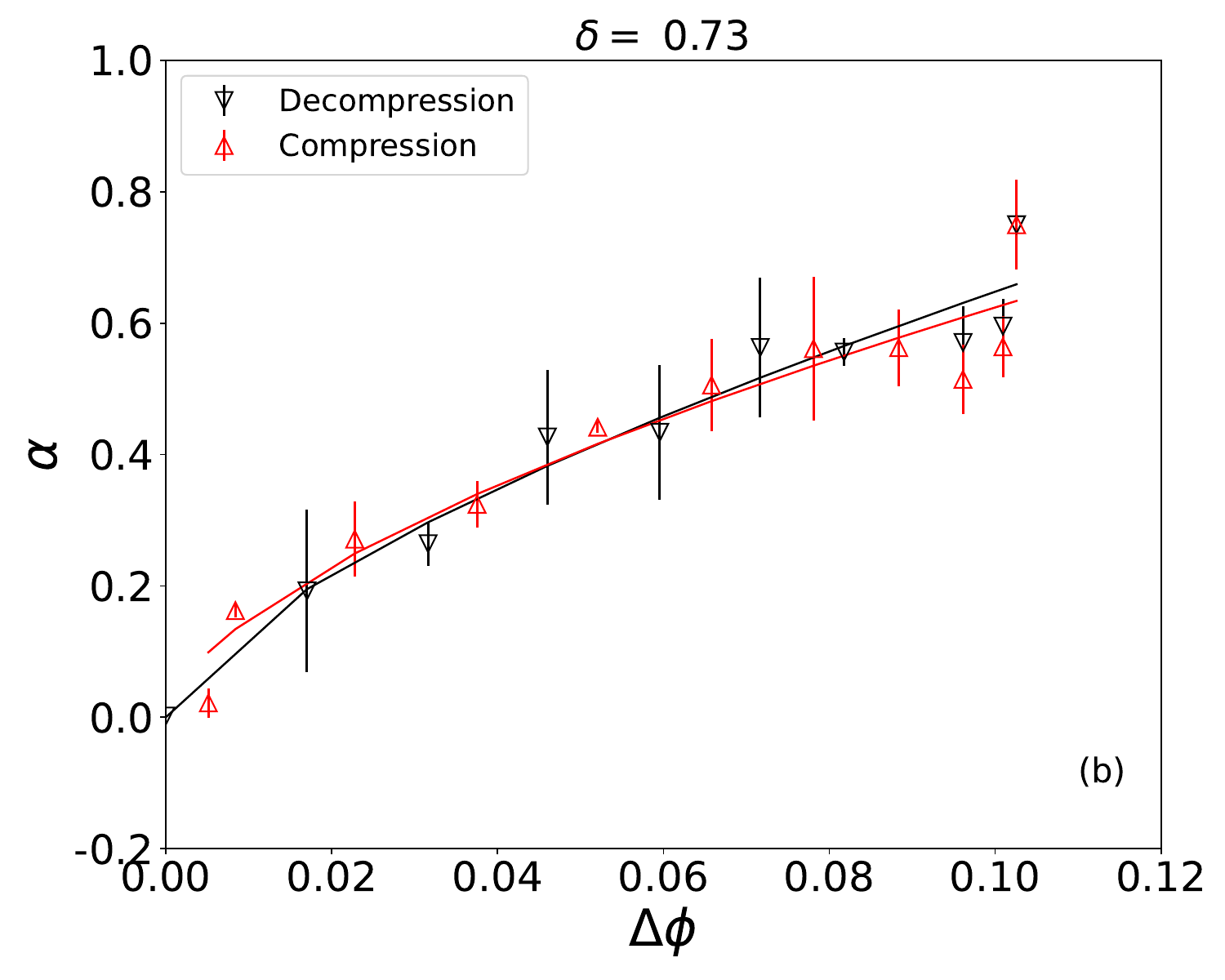}
    \caption{Low-$\omega$ exponent $\alpha$ as a function of $\Delta\phi = \phi - \phi_{J,uJ}$.
    The exponent $\alpha$ is shown during compression and decompression for (a) $\delta = 0$ 
    and (b) $\delta = 0.73$, extracted by fitting the low-$\omega$ region of the DOS in 
    Fig.~\ref{fig3} to  $D(\omega) \sim \omega^{\alpha}$. Continuous lines represent power-law 
    fits  $\alpha \approx \alpha_{0} + A \Delta\phi^{\nu_{\rm mono}}$  for monodisperse 
    and $\alpha \approx \alpha_{0} \Delta\phi^{\nu_{\rm bi}}$ for bidisperse ones, with extracted exponents:   
    $\nu_{\text{mono}} \approx 0.60$,  $\nu_{\text{bi}} \approx 0.62$ 
    during compression and  
    $\nu_{\text{mono}} \approx 0.59$,  $\nu_{\text{bi}} \approx 0.67$
    during decompression.  Error bars represent 
    the variance from the mean across three independent realizations, quantifying statistical 
    uncertainty.} 
	\label{fig4}
\end{figure}

The mechanical response of the packings under compression and decompression can be characterized by the density of states (DOS), \jp{which quantifies the distribution of vibrational modes as a function of frequency. It is defined as $g(\omega) = \frac{1}{N} \sum_{i=1}^{N} \delta(\omega - \omega_i)$, where $\omega_i$ is the frequency of the i-th vibrational mode, typically obtained by diagonalizing the dynamical matrix $\mathbf{D}$, see Sec.~IV of Ref~\cite{SupplementalMaterial}}. The DOS offers insight into the evolution of vibrational modes with changing packing density. Fig.~\ref{fig3} shows the DOS for monodisperse and bidisperse packings at three key states: near jamming (1), at $\phi_{\max} = 0.95$ (2), and near unjamming (3). Several intermediate cases are also included for comparison. During compression, typical van Hove singularities emerge at high frequencies near the jamming density in the monodisperse packing, as indicated by the open blue $\triangle$ in Fig.~\ref{fig3} (a). This observation aligns with the polycrystalline structure depicted in Fig.~\ref{fig2} (1) (left panel).
In the bidisperse system, prominent van Hove singularities are not observed due to the intrinsic disorder introduced by the different particle sizes, see Fig.~\ref{fig3} (b) and its 
corresponding configuration in Fig.~\ref{fig2} (1) (right panel). With further compression, the van-Hove singularities in the monodisperse case become more pronounced and broadened. Whereas in the bidisperse packing, two similar peaks appear at high $\omega$. The more pronounced case can be seen at $\phi_{\max}$ in Fig.~\ref{fig3}. These features indicate increased order after compression, as shown in Fig.~\ref{fig2} (2) and Fig.~\ref{fig0}.
A key feature of the DOS near jamming is the plateau at low $\omega$ for bidisperse packing. This is known as the boson peak, characterized by an excess of vibrational modes \cite{ohern2003jamming, Hecke2009Jamming, Silbert2005Vibrations, Goodrich2014Solids, Tong2015From, Zhang2021Disorder}. The DOS for the polycrystalline packing does not display a low-$\omega$ plateau. However, it still deviates from the expected Debye scaling. This deviation suggests that the vibrational properties of the polycrystalline packing more closely resemble those of a disordered packing rather than a single crystal structure. As the system undergoes further compression, the DOS evolves (open red $\bigcirc$) and, at $\phi_{\rm max}$ (solid $\Diamond$), follows the well-known Debye law at low $\omega$ (black dashed line). This behaviour suggests that polycrystalline monodisperse and disordered bidisperse packings behave as continuum elastic mediums in certain limits. Therefore, the disordered bidisperse structure shifts from behaving like disordered solids at low compression to resembling crystalline solid behaviour at high compression, see Fig.~\ref{fig3}. This transformation is particularly pronounced at $\phi_{\rm max}$, where the emergence of peaks at specified values of $\omega$ (see Fig.~\ref{fig3}), similar to those in the monodisperse case, suggests the development of some degree of local order. This observation is consistent with findings in 3D bidisperse packings, as noted in \cite{petit2023structural, clarke1993structural, klumov2014structural, hanifpour2015structural}, where local order appears at the jamming density and becomes increasingly prominent with further compression.

During decompression, the DOS for both systems evolves (solid $\bigcirc$), exhibiting behaviour similar to that observed during compression near unjamming. Interestingly, despite the distinct local structural features characterised by nonaffine displacements~\cite{Ganguly2013} (see Fig.~\ref{fig2}) at $\phi_J$ and $\phi_{uJ}$ values observed for monodisperse and bidisperse packings near the jamming and unjamming states, their corresponding density of states remains strikingly similar at these points (see Fig.~\ref{fig3}). Even the intermediate DOS remains similar at comparable distances from the jamming/unjamming densities. A difference in the DOS is expected as the packing structure evolves during a complete cycle of compression and decompression. This structural change has been shown to affect the jamming density in previous work \cite{kumar2016memory}.  

To better understand the evolution of the DOS during compression-decompression, particularly at low frequencies, we fitted the low-$\omega$ region in Fig.~\ref{fig3} with a power law, $D(\omega) \sim \omega^{\alpha}$. The extracted exponent is shown in Fig.~\ref{fig4}. For monodisperse packings, 
$\alpha$ remains low but nonzero near the jamming/unjamming density, see Fig.~\ref{fig4} (a). As 
$\Delta\phi$ increases, $\alpha$ undergoes a discontinuous change from the jamming/unjamming state, then 
rises continuously until reaching $\alpha = 1$ at high $\Delta\phi$, consistent with Debye scaling. In 
contrast, for bidisperse packings, $\alpha$ transitions smoothly from zero—indicating a plateau—near the 
jamming/unjamming density to $\alpha \approx 0.8$ at high $\Delta\phi$, approaching Debye scaling, see 
Fig.~\ref{fig4} (b). The relationship between the low-frequency exponent $\alpha$ and the 
distance from the jamming or unjamming point, $\Delta\phi = \phi - \phi_{J,uJ}$, follows a power-law 
trend: $\alpha \approx \alpha_0 + A\, \Delta\phi^{\nu_{\rm mono}}$ for monodisperse packings and 
$\alpha \approx \alpha_0\, \Delta\phi^{\nu_{\rm bi}}$ for bidisperse ones. The fitted exponents 
are $\nu_{\rm mono} \approx 0.60$ and $\nu_{\rm bi} \approx 0.62$ during compression, and 
$\nu_{\rm mono} \approx 0.59$, $\nu_{\rm bi} \approx 0.67$ during decompression.

The nonzero $\alpha$ near the jamming/unjamming transition in polycrystalline packings confirms the absence of a low-$\omega$ plateau in the DOS while still deviating from Debye scaling. The discontinuity in $\alpha$ may be related to rearrangements near jamming/unjamming, as indicated by the change in mean contact number (from 0 to greater than 5) shown in Fig.~\ref{fig0} (c)
and localised nonaffine displacements of the particles around the grain boundaries (left panel of Fig.~\ref{fig2}). In contrast, for bidisperse disordered packings, the smooth evolution of the DOS and $\alpha$ is accompanied by apparent differences in mean contact numbers between large and small particles, which arise due to variations in the rattler fraction rather than distinctions in the contact network. The absence of the plateau and the discontinuity in polycrystalline packings may also stem from the strain rate used ($\dot{\epsilon} \sim10^{-4}$), far higher than the slower process performed in \cite{Goodrich2014Solids}, where a plateau is observed for polycrystals. Therefore, this proves to be a more nuanced problem that requires additional simulations at varying strain rates to capture subtle changes in DOS in the low $\omega$ regime as the system undergoes fast or quasistatic compression-decompression cycles.

 \begin{figure*}[t]
    \centering \includegraphics[scale=0.55]{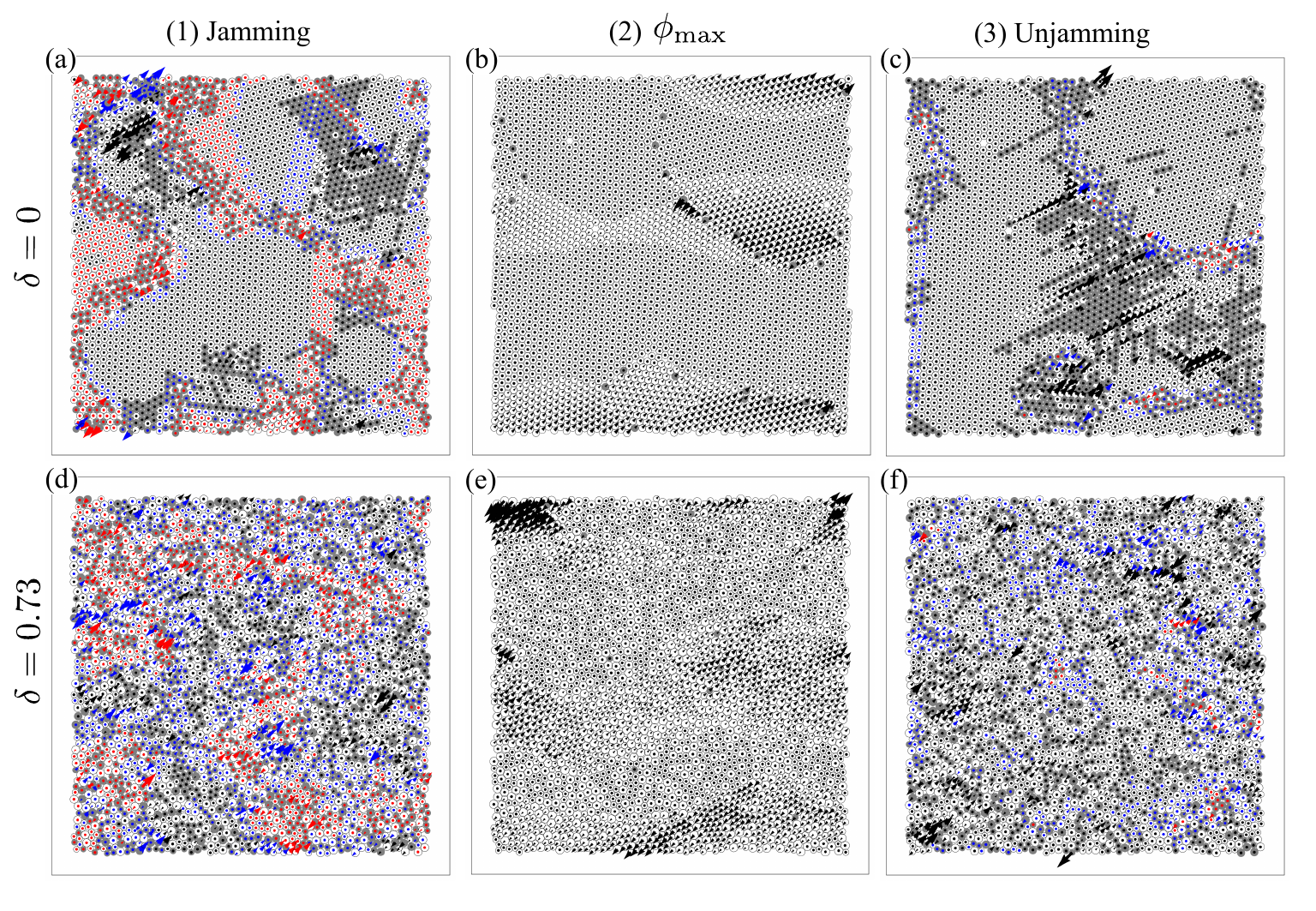}
    \caption{Eigenvectors for $\omega=0.05$. (Top panel) Eigenvectors of the monodisperse 
    and (Bottom panel) bidisperse systems associated with packing fractions near jamming (1), 
    at $\phi_{\rm max}$ (2) and near unjamming (3). The displacement fields marked with red, blue, and 
    black arrows indicate large ($\chi_{i} \geq 5$), intermediate ($1 \leq \chi_{i} < 5$), 
    and low ($\chi_{i} < 1$) nonaffine particle values respectively. Gray disks identify particles with less than four contacts (rattlers).
    Note the wave-like behavior of the particles at $\phi_{\rm max}$ consistent with the elastic 
    nature of the system. Jamming and unjamming give rise to localized soft modes, which drive the deviation from Debye scaling in monodisperse packings and the emergence of a plateau in bidisperse packings, as observed in Fig.~\ref{fig3}.} 
	\label{fig5}
\end{figure*}

The evolution in the vibrational density of states suggests an underlying connection between structural rearrangements and the evolution of vibrational modes. To further investigate this relationship, we analyze the changes in the packing structure by computing the contact orientational order (COR), $\mathcal{Q}_{6}$, which quantifies the formation of potential six-fold 
lattice structures (see its definition in Sec.~V in \cite{SupplementalMaterial}). This variable avoids the ambiguities of the bond orientational order (BOR) and is 
more sensitive to sudden changes in contact numbers or other structural features compared to the structure factor, see Ref.~\cite{petit2023structural}. Figs.~\ref{fig0} (e)-(f) display the evolution of $\mathcal{Q}_{6}$ for monodisperse and bidisperse packings. Monodisperse packings exhibit variations in $\mathcal{Q}_{6}$ after jamming due to particle rearrangement, stabilizing under further compression, see Figs.~\ref{fig0} (e). Notably, $\mathcal{Q}_{6}$ at $\phi_{\rm max}$ remains below 1, indicating that the system is not fully ordered and deviates significantly from the ideal value of $\mathcal{Q}_{6} = 1$ expected for perfect hexagonal packing \cite{Tong2015From}. During decompression, $\mathcal{Q}_{6}$ slightly increases, suggesting some organization due to stored energy before dropping near unjamming (see Fig.~\ref{fig2}).
For bidisperse packings, we quantify local $\mathcal{Q}_{6}$ values between Large-Large (LL), Small-Small (SS), and Small-Large (SL) particles, each defined in \cite{SupplementalMaterial}. Fig.~\ref{fig0} (f) illustrates that SL contacts predominantly govern the evolving structure, while LL and SS contacts are slightly reduced near $\hat{\phi} = 0$. This reduction occurs because particles of the same size experience slight overlap at  $\phi_{\rm max}$, leading to distortions that introduce disorder into the packing. Interestingly, all partial $\mathcal{Q}_{6}$ 
values near jamming (1) and unjamming (3) are identical, see the arrows in Fig.~\ref{fig0} (f). 
This suggests that while the bidisperse packing structure evolves during compressions,  configurations are restored upon decompression, such that jamming and unjamming states exhibit similar values for the structural quantifiers like $\mathcal{Q}_{6}$. Identifying jamming and unjamming configurations in monodisperse packings in Fig.~\ref{fig0} (c) is more challenging. Variations in $\mathcal{Q}_{6}$ near jamming/unjamming further support the idea that rearrangements may be responsible for the absence of the plateau and the $\alpha$ discontinuity in polycrystalline systems. The arrows in panels (1)-(3) of Fig.~\ref{fig0} (c) highlight similar local $\mathcal{Q}_{6}$ values, consistent with the structure factor in \cite{SupplementalMaterial}. Therefore, the preserved packing structure in both monodisperse and bidisperse systems near jamming/unjamming likely results in comparable vibrational and mechanical behaviour. This is confirmed in Fig.~\ref{fig3} and reinforced by the low-$\omega$ DOS exponent in Fig.~\ref{fig4}, which remains similar near jamming and unjamming, indicating a consistent structural influence on vibrational properties.

The local configurations near jamming and unjamming can also be explored by quantifying the nonaffine motion of particles during compression-decompression and 
examining its influence on the low-$\omega$ behavior of the DOS in Fig.~\ref{fig3}. Figs.~\ref{fig0} (g)-(h) depict the evolution of cumulative nonaffine displacements ($\mathcal{X}_{q}$) \cite{Ganguly2013,Ganguly2015} as a function of $\hat{\phi}$ for both monodisperse and bidisperse packings, with $q \in$ [L,S] (see Sec.~VI of \cite{SupplementalMaterial}). $\mathcal{X}_{q}$ consistently decreases during compression showing a sudden drop near jamming for monodisperse packing, whereas for bidisperse packing is continuous. The sharp decrease in $\mathcal{X}_{q}$ near $\hat{\phi} = 0$ in both systems indicates that an increase in density close to $\phi_{\rm max}$ significantly restricts the movement of nonaffine particles. Instead, the particles undergo affine displacements, indicating that the systems follow an elastic behaviour. During decompression, the nonaffine motions increase again, but are more continuous, showing fewer discontinuous jumps. Since discontinuous changes in $\mathcal{X}$ signal significant particle rearrangements~\cite{Ganguly2017}, the absence of these discontinuities during decompression suggests that the overall packing structure at $\phi_{\mathrm{max}}$ is preserved. These findings align with the consistent $\mathcal{Q}_{6}$ values in Figs.~\ref{fig0} (e)-(f) across jamming and unjamming states and with 
the $S(q)$ data presented in \cite{SupplementalMaterial}.

Nonaffine displacements are analyzed near jamming, at $\phi_{\rm max}$, and near unjamming to assess their impact on the low-$\omega$ vibrational modes in polycrystalline and disordered packings. Eigenvectors for $\omega = 0.05$, which lie within the plateau region or deviate from Debye’s law in the DOS, are extracted and shown in Fig.~\ref{fig5}. Each eigenvector is color-coded based on the magnitude of nonaffine displacement of each particle $(\chi_{i})$: red for high $(\chi_{i} > 5)$, blue for intermediate $(1 \leq \chi_{i} < 5)$, and black for low $(\chi_{i} < 1)$. At $\phi_{\rm max}$, eigenvectors in polycrystalline monodisperse and disordered bidisperse packings display wave-like behavior with minimal nonaffine motion, reflecting macroscopic solid behavior consistent with Debye’s law (see Fig.~\ref{fig5} (b)/(e)). In contrast, near jamming and unjamming, several localized modes emerge with varying nonaffine displacements, particularly in regions prone to rearrangements (see Fig.~\ref{fig5}(a)/(c) for monodisperse and (d)/(f) for bidisperse packings). While Fig.~\ref{fig5} (a)/(c) and (d)/(f) reveal differences in the local structures in the eigenvectors near jamming and unjamming, the low-$\omega$ DOS and the exponent $\alpha$ remain similar.
In addition, rattlers (colored gray) appear in distinct clustered structures in monodisperse systems near jamming and unjamming, whereas in bidisperse packings, they are more dispersed and localized, further influencing the vibrational properties of each system
in a distinct manner.

The results presented in this study provide insights into the similarities in the vibrational behavior of polycrystalline and disordered granular solids during jamming and unjamming transitions. Using discrete element method simulations, the research investigates the vibrational density of states (DOS) near jamming and unjamming, emphasizing deviations from Debye scaling and the role of nonaffine displacements. A key finding is that while bidisperse packings exhibit a low-$\omega$ plateau, polycrystalline packings do not, yet both systems deviate from Debye’s law. The DOS for bidisperse packing smoothly evolves from zero (plateau) to near one (Debye scaling) which is captured by the low-$\omega$ exponent $\alpha$. The presence of two particle sizes promotes a more uniform local structure, as reflected in the orientation order parameter $Q_6$ in Fig.~\ref{fig0} (f) and the local nonaffine displacement shown in the right panel of Fig.~\ref{fig2}. In contrast, for monodisperse packing, $\alpha$ exponent exhibits a non-smooth behaviour near jamming/unjamming. Unlike in the bidisperse system, the polycrystalline structure exhibits nonaffine displacements (left panel of Fig.\ref{fig2}) that are more localized around grain boundaries, along with large rearrangements marked by a sudden change in $Q_6$ (Fig.~\ref{fig0} (e)). This may explain the absence of the plateau in this system. 
We found that $\alpha$ (see Fig.~\ref{fig4}) remained consistent across compression and decompression for similar values of packing fractions. Despite structural modifications occurring throughout the process, no significant history dependence is observed in the DOS, suggesting that vibrational properties are preserved across jamming and unjamming transitions.
The study also highlights the role of structural features, such as the evolution of the contact network, in shaping the vibrational response. Analysis of nonaffine displacements and contact orientational order reveals that while structural changes occur during compression, features of the configurations contributing to the macro-scale quantifiers are largely restored upon decompression, contributing to the vibrational similarities between jamming and unjamming states.

These findings point toward a connection between microstructural features—such as contact network rearrangements, orientational order, and nonaffine displacements—and the vibrational response of granular materials during jamming and unjamming. While the results highlight distinct behaviors between polycrystalline and disordered packings, particularly in the evolution of the low-frequency density of states, further work is needed to generalize these observations. Nonetheless, the trends observed here provide a promising basis for future investigations into the role of microstructure in governing mechanical stability in disordered solids.
Future work will focus on extending these findings to a broader range of size ratios and small particle concentrations, with particular emphasis on the first and second jamming transitions in bidisperse packings. In these systems, a discontinuous transition has been observed, separating jammed states dominated by large particles from those involving both large and small particles \cite{petit2020additional}. This extension of the current study will deepen our understanding of the complex interplay between particle size distribution and the jamming behavior in granular materials.

\begin{acknowledgments}
We thank Till Kranz and Matthias Schr\"{o}ter, for proofreading, 
fruitful discussions, and providing constructive criticism about 
results and the paper. This work was supported by the German Academic Exchange Service (DAAD)
under grant n$^{o}$ 57424730. SG acknowledges support from Deutsche
Forschungsgemeinschaft through grant FU 309/11-1.
\end{acknowledgments}

\bibliography{ref}
\bibliographystyle{apsrev4-2} 

\end{document}


\begin{center}{\LARGE Supplementary Material: 
\\
\bigskip
Vibrational similarities in jamming-unjamming of polycrystalline and disordered granular packings}
\linebreak[1]

Juan C. Petit\footnote{Juan.Petit@dlr.de},
Saswati Ganguly, 
and Matthias Sperl\\ 
Institut f\"{u}r Materialphysik im Weltraum, Deutsches Zentrum f\"{u}r Luft- und Raumfahrt (DLR), 51170 K\"{o}ln, Germany\\
Institut f\"{u}r Theoretische Physik, Universit\"{a}t zu K\"{o}ln, 50937 K\"{o}ln, Germany

\end{center}

\section{Technique and procedure of the simulation}
\label{SecI}

We perform molecular dynamic simulations of monodisperse and bidisperse frictionless soft-disk
jammed packings in 2D without gravity \cite{cundall1979discrete, weinhart2020fast, petit2022bulk}. Zero gravity is employed to investigate how the DOS evolves during the development of jamming and unjamming structures, a process that is not possible to achieve under the influence of gravity.
$N = 3000$ particles form the packings 
where their evolution is numerically solving Newton's equation. The bidisperse packing has a number of large, $N_{\mathrm L}$, and small, $N_{\mathrm S}$, particles 
with radius $r_{\mathrm L}$ and $r_{\mathrm S}$. The size ratio is $\delta = r_{\mathrm{S}}/r_{\mathrm{L}} = 0.73$ with a concentration of small particles of 
$X_{\mathrm S} = N_{\mathrm S} \delta^{2} / (N_{\mathrm L} + N_{\mathrm S} \delta^{2}) = 0.5$. Therefore, $\delta = 0$ represents the monodisperse packing. $\delta = 0.73$ and $X_{\mathrm{S}} = 0.5$ are chosen to prevent long-range order \cite{majmudar2007jamming, ohern2003jamming, morse2016geometric}, ensuring the 
emergence of local structures that reflect the real-world granular behaviors.

The linear spring-dashpot model, ${\it \bf f}^n_{ij}=f^n_{ij} \hat{\bf n} = ( k_{n} \alpha_c + \gamma_{n} \dot\alpha_c) \hat{\bf n}$, is used for the contact between particles 
\cite{cundall1979discrete, luding2008cohesive, petit2022bulk}, where $k_{n}$ is the contact spring stiffness, $\gamma_{n}$ is the contact damping coefficient, $\alpha_c$ is the contact overlap and $\dot\alpha_c$ is the relative velocity in the normal direction $\hat{\bf n}$. A background dissipation force is imposed 
on each particle velocity, with constant dissipation, to damp out the kinetic energy of the particles.
We use as a unit of length and time, $r_{\rm L}$ and $t = \sqrt{m_{\rm L}/k}$ respectively, where 
$m_{\rm L}$ is the mass of the large particles and $k_{n}$ is the stiffness constant.

The initial configuration of the packing is such that disks are uniformly generated at random in a 2D box with uniform random velocities and with an initial packing fraction of $\phi_{\mathrm{ini}} = 0.3$. Large overlaps lead to an initial peak in kinetic energy, but this is quickly damped by the background medium and collisions. Low-density systems with high kinetic energy contribute to the rapid randomization of particles. The granular gas is then isotropically compressed to approach 
an initial direction-independent configuration with the target packing fraction $\phi_0 = 0.8$ slightly below the jamming density at $\phi_J$, see Ref.~\cite{petit2022bulk} for further details. The system then undergoes a relaxation process to eliminate any remaining kinetic energy. Once relaxation is complete, isotropic compression begins and continues until the packing fraction reaches the maximum value, $\phi_{\mathrm{max}} = 0.95$. Then, the isotropic decompression 
process proceeds at the same rate as the compression until $\phi_{0}$ is reached again.
The mean volumetric strain rate over time, $\langle\dot{\varepsilon}_{\rm v}\rangle$, during the compression-decompression process in this study is approximately one hundred times greater than in previous work \cite{kumar2016memory}, reaching values around $\langle\dot{\varepsilon}_{\rm v}\rangle \sim 10^{-4}$. Despite this accelerated rate, the protocol successfully generates polycrystalline monodisperse and disordered bidisperse packings. Moreover, the protocol maintains a stable energy ratio—kinetic to potential energy—of approximately $10^{-4}$ near the jamming and unjamming transitions, ensuring that the packings remain in mechanical equilibrium throughout the process.

The compression (decompression) protocol consists in simultaneously moving all periodic boundaries inward (or outward). The motion of the periodic boundaries is modeled using a cosine function. For example, the vertical wall position (or height) is given by 
\begin{equation}
z(t) = z_{f} + \frac{z_{0}-z_{f}}{2}(1 + \mathrm{cos}(2\pi\,(t-t_0)/T)),
\label{ecu0} 
\end{equation}  
where the strain is defined as $\varepsilon_{zz}(t) = 1 - z(t)/z_{0}$, with $z_{0}$ and $z_{f}$ the initial and maximum vertical dimensions of the box at zero and maximum strain, respectively, and $T$ is the total simualtion time. During the compression and decompression paths, the strain rate is given by 
$\dot{\varepsilon}_{\rm v} = \dot{\varepsilon}_{xx} 
+ \dot{\varepsilon}_{zz} \propto (1 - z_f/z_0)\, \mathrm{sin}(2\pi (t - t_0)/T)/T$. The time-offset, $t_0$, ensures a smooth start-up and finish of the boundary motion
via a co-sinusoidal function, avoiding shocks and inertia effects. 
At the conclusion of the simulation protocol, the jamming density of each bidisperse 
packing is determined from the decompression branch. This approach is preferred as 
the decompression branch is less sensitive to deformation rates, as noted 
by Ref.~\cite{goncu2010constitutive}.


\section{Extracting the jamming and unjamming densities}
\label{SecII}

\begin{figure}[t]
    
    \centering \includegraphics[scale=0.28]{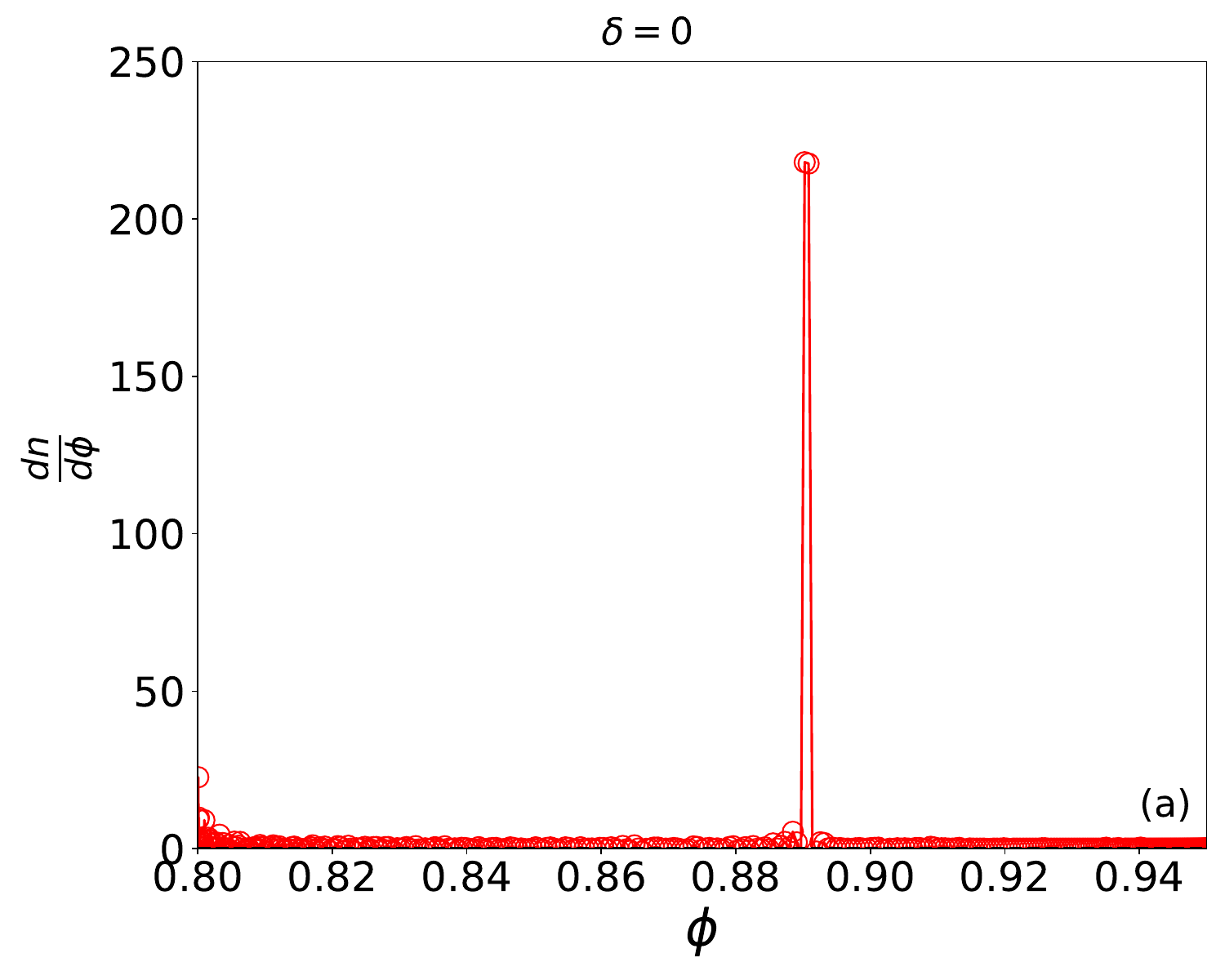}
    \centering \includegraphics[scale=0.3]{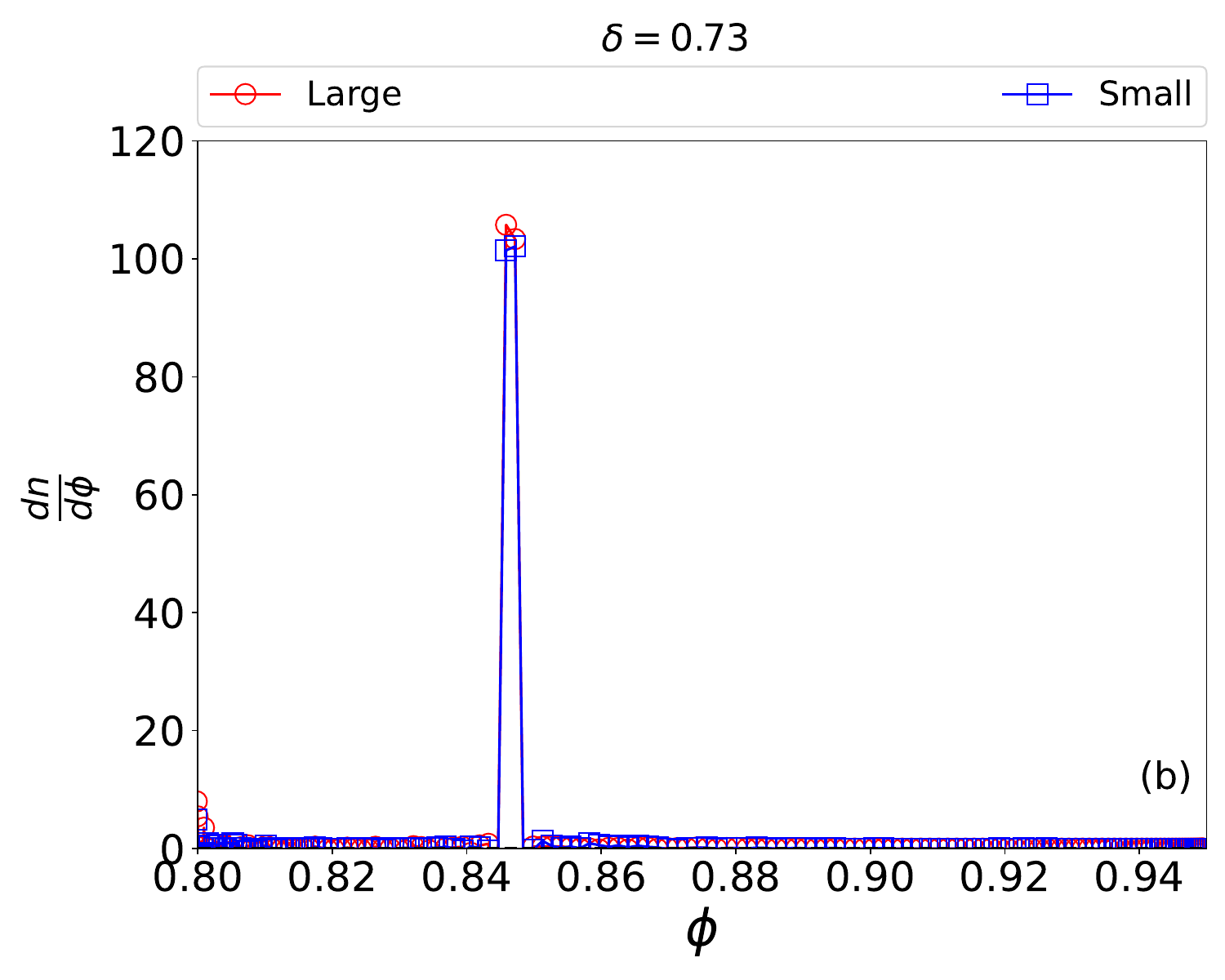}

    \caption{Derivative in the fraction of particles as a function of $\phi$ for (a) monodisperse and (b) bidisperse packings along the decompression path. The peak shows the maximum number of particles jammed at $\phi_{uJ}$. 
    Similar results are obtained along the compression path.}
	\label{dndphi}
\end{figure}

The derivative in the fraction of particles (n) of monodisperse and bidisperse packings are used 
to extract the values of the jamming ($\phi_J$) and unjamming ($\phi_{uJ}$) densities, i.e., the 
value at which $n$ show a discontinuous behavior. Such behavior represents a jamming or unjamming 
transition of the granular packing upon compression or decompression. Mathematically speaking, the 
derivative is determined using the five-point finite difference method over the data shown in Fig.~1 (a-b) of the 
paper. This method gives a precise value of $\phi_J$ and $\phi_{uJ}$ for both packings. 
Fig.~\ref{dndphi} shows the derivative for (a) monodisperse and (b) bidisperse packings as a function 
of $\phi$ along the decompression path. The method shows a characteristic peak (maximum derivative) at a 
value consistent with $\phi_{uJ}$. A similar peak is obtained during compression (not shown).
Note that for $\delta = 0.73$ and $X_{\mathrm S} = 0.5$, the peak for both large and small particles 
shows the same packing density of $\phi_{uJ}\approx 0.84$ (with $\phi_{J}\approx 0.83$). This means that both 
types of particles simultaneously unjammed (jammed). For the monodisperse case, we obtained 
$\phi_J \approx 0.87$ and $\phi_{uJ} \approx 0.88$, respectively.

\section{Structure factor quantification}
\label{SecIII}

We calculate the partial structure factors, $S(q)$, near jamming, at $\phi_{\mathrm{max}}$, and near unjamming. 
This analysis enables us to investigate the structural contributions and structural changes associated with 
each configuration type (in the case of bidisperse packings) within both jammed and unjammed states. 
A general definition of the partial $S(q)$ is

\begin{equation}
\begin{split}
S_{\mathrm{\nu\beta}}(q)\ &= \frac{1}{N} \sum_{i=1}^{N_{\mathrm{\nu}}} \sum_{j=1}^{N_{\mathrm{\beta}}}
							   \mathrm{cos}(\mathbf{q}\cdot \mathbf{r}^{i}_{\mathrm{\nu}}) \mathrm{cos}(\mathbf{q}\cdot \mathbf{r}^{j}_{\mathrm{\beta}})\\  	
							 & + \frac{1}{N} \sum_{i=1}^{N_{\mathrm{\nu}}} \sum_{j=1}^{N_{\mathrm{\beta}}} 
							   \mathrm{sin}(\mathbf{q}\cdot \mathbf{r}^{i}_{\mathrm{\nu}}) \mathrm{sin}(\mathbf{q}\cdot \mathbf{r}^{j}_{\mathrm{\beta}})  
\end{split}
\label{ecu2}
\end{equation}

\begin{figure}[t]
    
    \centering \includegraphics[scale=0.28]{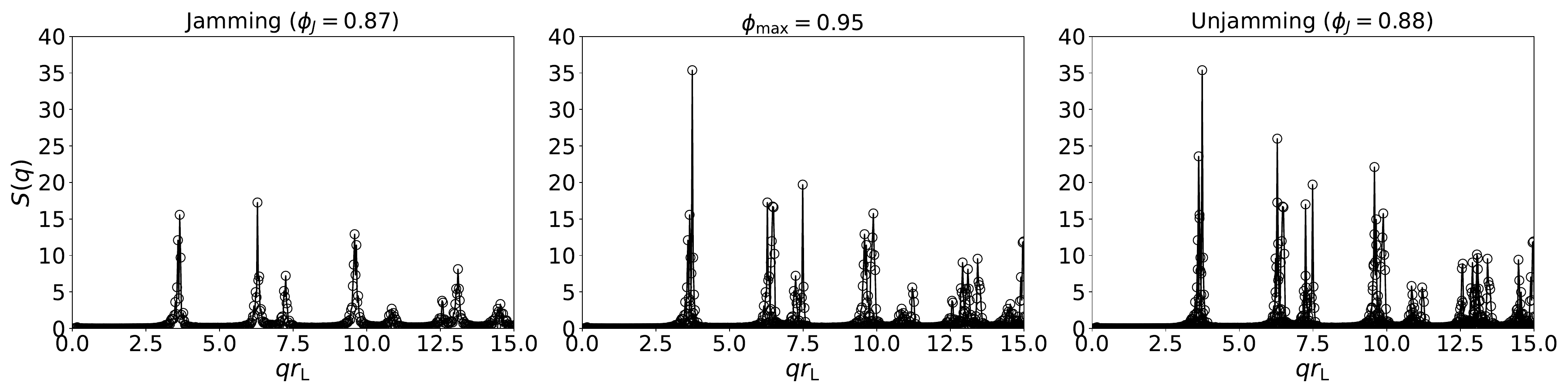}
    \centering \includegraphics[scale=0.28]{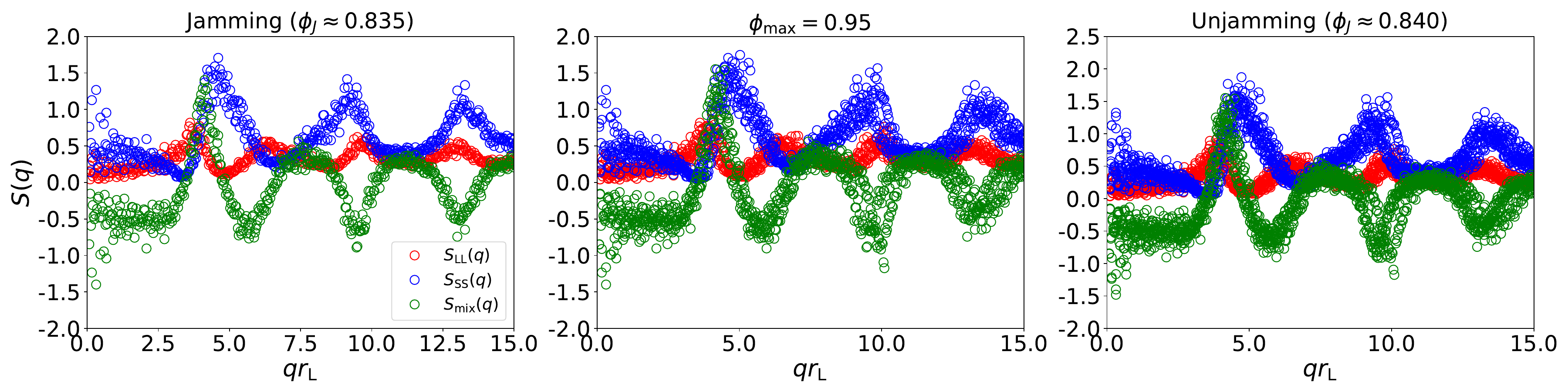}

    \caption{The structure factor, $S(q)$, is shown for both monodisperse (top) and bidisperse 
    packings (bottom) across three states: near jamming, at maximum packing density $\phi_{\mathrm{max}}$, 
    and near unjamming. Notably, minimal changes are observed in the packing structure across these states, 
    indicating a consistent organization of particles despite variations in packing density. }
	\label{structFact}
\end{figure}

\noindent  where $\nu$,$\beta \in \{\mathrm{L},\mathrm{S}\}$ and the sum runs over all $\nu$ 
and $\beta$ particles. Thus, the total structure factor $S(q)$ can be decomposed into contributions: 
$S_{\mathrm{ LL }}(q)$, $S_{\mathrm{ SS }}(q)$, $S_{\mathrm{ LS }}(q)$, and $S_{\mathrm{ SL }}(q)$. 
By symmetry, we have $S_{\mathrm{ LS }}(q) = S_{\mathrm{ SL }}(q)$.  For simplicity, we label this 
shared term as $S_{\mathrm{mix}}(q)$, where “mix” emphasizes the equivalence between 
$S_{\mathrm{LS}}(q)$ and $S_{\mathrm{SL}}(q)$. Therefore, the total structure factor can be 
expressed as $S(q) = S_{\mathrm{ LL }}(q) + S_{\mathrm{ SS }}(q) + 2S_{\mathrm{mix}}(q)$.
In monodisperse packings, $S(q) = S_{\mathrm{LL}}(q)$ is only calculated. Fig.~\ref{structFact} 
illustrates the structure factors for both monodisperse and bidisperse packings near jamming, 
at $\phi_{\mathrm{max}}$, and near unjamming. Near jamming, we observe multiple peaks in $S(q)$ 
for the monodisperse packing, reflecting its polycrystalline nature. In contrast, the bidisperse 
mixture exhibits a partial $S(q)$ with broader peaks, indicative of a disordered structure with 
local ordering. We interpret this as a result of the bidisperse packing being locally ordered 
but globally disordered. Further compression increases the density of both systems, as seen in 
the development of new peaks and the growth of existing peaks, particularly at 
$\phi_{\mathrm{max}}$ in both monodisperse and bidisperse packings.
Upon decompression, only minimal structural changes arise from particle rearrangements, yet 
the jamming and unjamming states retain similar structural features. This suggests that they 
are structurally comparable, although the unjamming state may exhibit slightly higher 
local particle densities.


\section{Density of states (DOS)}
\label{SecIV}

We present the density of states (DOS) throughout the compression-decompression process, with a focus on the DOS near jamming and unjamming. It is defined as $g(\omega) = \frac{1}{N} \sum_{i=1}^{N} \delta(\omega - \omega_i)$, quantifying the number of vibrational modes per unit frequency.
This comparison emphasizes the vibrational behavior of polycrystalline monodisperse and disordered bidisperse packings in relation to the structural features discussed earlier.

The DOS is calculated by diagonalizing the dynamical matrix, $\mathbf{D}$, where
the obtained eigenvalues are related to the frequencies of the system and the eigenvectors with the vibrational modes of the particles. Off-diagonal elements of $\mathbf{D}$ from different particles $i \neq j$ are calculated by 

\begin{equation}
    D_{ij} = - k \frac{\vec{r}_{ij} \otimes \vec{r}_{ij} }{r^{2}_{ij}} +  \frac{f_{ij}}{r_{ij}} \bigg( I - \frac{\vec{r}_{ij} \otimes \vec{r}_{ij} }{r^{2}_{ij}} \bigg),
    \label{ecu1}
\end{equation}

\noindent where $\vec{r}_{ij}$ and $f_{ij}$ are the distance vector and force magnitude between 
particle $i$ and $j$, respectively. $I$ denotes the unit tensor. The diagonal elements from same 
particle $i = j$ are determined by $D_{ii} = - \sum_{j \in \Gamma_{i}} D_{ij}$, where $\Gamma_{i}$
specifies the number of $j$ particles in contact with particle $i$. We have also considered that 
the particle mass is one, see Ref.~\cite{hara2023heterogeneity}.


\section{Definition of the contact orientational order (COR), $\mathcal{Q}_{6}$.}
\label{SecV}

To understand the evolution of the packing structure, we 
calculate the contact orientational order (COR) defined as $\mathcal{Q}_{6} = \frac{1}{N}\sum_{j = 1}^{N} \mathcal{Q}_{6j}$
where $\mathcal{Q}_{6j} = \big| \frac{1}{N_{c}^{j}} \sum_{k = 1}^{N_{c}^{j}} e^{6i\theta_{kj}}\big|$, which 
quantifies the formation of potential six-fold lattice structures within the packing. 
Here, $N_{c}^{j}$ represents the number of contacts for the $j$th particle, and $\theta_{kj}$ 
is the contact angle between particles $k$ and $j$,  measured with respect to the x-axis. This approach, 
introduced in Ref.~\cite{petit2023structural}, avoids neighborhood ambiguities inherent 
in the bond orientational order (BOR) metric, see Refs.~\cite{steinhardt1983bond, martin2008influence, duff2007shear, 
fortune1995voronoi, kansal2002diversity, mickel2013shortcomings, haeberle2019distinguishing}, 
and is sensitive to sudden changes in contact number or other structural features. 
Rather than using a separate detection method to identify nearest neighbors, COR is defined 
directly by particle contacts. This means that, for each particle $i$, its neighbors 
are implicitly defined by contact. According to the definition of $\mathcal{Q}_{6}$, 
in the absence of a jammed structure (i.e., before jamming), $\mathcal{Q}_{6}$ 
should ideally be zero. However, isolated clusters of contacting particles may 
yield low, nonzero $\mathcal{Q}_{6}$ values. Unlike prior definitions where local bond orientational order in highly 
amorphous packings is expected to vanish at jamming, COR allows for a finite, 
low $\mathcal{Q}_{6}$ value at the onset of jamming in amorphous packings. On the other hand, 
high values of $\mathcal{Q}_{6}$, particularly those close to one, indicate an ordered hexagonal
packing structure. Notably, for dense packings where  $\phi \gg \phi_{J,uJ}$, we expect 
BOR and COR converge, as BOR’s detection method will naturally identify 
those particles in contact with particle $i$ as its nearest neighbors. To explicitly understand 
the role of the structure and make the analysis clearer, we split $\mathcal{Q}_6$ on the types of 
particle contacts: Large-Large (LL), Small-Small (SS) and Large-Small or Small-Large (SL/LS), 
where their definition are 

\begin{equation}
    Q_6^{\mathrm{LL}} = \frac{1}{N} \sum_{j \in L}^{N^{\mathrm{LL}}} \bigg| \frac{1}{N_c^{j\mathrm{L}}} \sum_{k \in \mathrm{L}}^{N_c^{j\mathrm{L}}} e^{6i\theta_{jk}} \bigg|,
    \label{eq4.1}
\end{equation}

\begin{equation}
    Q_6^{\mathrm{SS}} = \frac{1}{N} \sum_{j \in \mathrm{S}}^{N^{\mathrm{SS}}} \bigg| \frac{1}{N_c^{j\mathrm{S}}} \sum_{k \in \mathrm{S}}^{N_c^{j\mathrm{S}}} e^{6i\theta_{jk}} \bigg|,
    \label{eq4.2}
\end{equation}

\begin{equation}
    Q_6^{\mathrm{LS}/\mathrm{LS}} = \frac{1}{N} \sum_{j \in \mathrm{L}}^{N^{\mathrm{LS}}} \bigg| \frac{1}{N_c^{j\mathrm{S}}} \sum_{k \in \mathrm{S}}^{N_c^{j\mathrm{S}}} e^{6i\theta_{jk}} \bigg|.
    \label{eq4.3}
\end{equation}

Figs.~3 (c)-(d) in the paper shows the evolution of $Q_6^{\mathrm{LL}}$, $Q_6^{\mathrm{SS}}$
and $Q_6^{\mathrm{LS}/\mathrm{LS}}$ as a function of packing fraction. 


\section{Cumulative nonaffine displacements ($\mathcal{X}$)}
\label{SecVI}

The definition of the nonaffine parameter per particle $\chi_{i}$ and the cumulative nonaffine parameter $\mathcal {X}$ are taken from previously established definitions of this quantity in references \cite{Ganguly2013, Ganguly2015}. 

The nonaffine parameter $\chi_{i}$ provides a geometric measure of local structural deviations from homogeneous, 
system-wide deformations due to applied volume or shear strains. Defined over a chosen coarse-graining 
neighborhood volume $\Omega$ around a particle $i$, the nonaffine parameter $\chi_{i}$ and 
the best-fit affine strain $\bm{\epsilon}_{i}$ are given by

\begin{eqnarray}
\chi_{i}&=&\mathsf{\Delta}\mathsf{P}\mathsf{\Delta}, \text{ with } \ \ {\mathsf P} =\mathsf{I}-\mathsf{RQ}, \label{def_chi}\\
\bm{\epsilon}_{i}&=&{\mathsf Q}\mathsf{\Delta}, \text{ where } \ \ {\mathsf Q} = ({\mathsf R}^{T}{\mathsf R})^{-1}{\mathsf R}^{T}\label{def_e}.
\end{eqnarray}
Here, $\chi_{i}$ and $\bm{\epsilon}_{i}$ are projections of the displacement fields $\mathsf{\Delta}$ using 
the projector $\mathsf{P}$, based on a reference configuration $\mathsf{R}$ \cite{Ganguly2013}. To 
compute $\chi_{i}$ for each particle in an N-particle configuration, the displacement fields $\mathsf{\Delta}$
within the neighborhood $\Omega$ are compared to the reference relative positions $\mathsf{R}$. For 
defect-free crystalline systems under deformation or thermal fluctuations, $\mathsf{R}$ is typically
chosen as the zero-temperature energy minimum configuration. In case of an ideal triangular lattice, the course graining volume $\Omega$ can be chosen as the nearest neighbour distance. For polycrystalline monodisperse or 
disordered bidisperse packings, we use the configuration at $\phi_{\rm max}$ as the reference, 
leveraging its mechanical stability to analyze microstructural changes during jamming
and unjamming in granular assemblies.
The parameter $\chi_{i}$, with units of $(\rm length)^{2}$, can be derived from the eigenvalues of 
the matrix $\mathsf{PCP}$, which itself is projected from the covariance matrix $\mathsf{C} = \langle \mathsf{\Delta}^{T} \mathsf{\Delta} \rangle$ \cite{Ganguly2013}. Prior work (e.g., Refs.\cite{Ganguly2015, Ganguly2017, PREpopli2019, Mitra2015a}) 
demonstrates that the eigenvectors of $\mathsf{PCP}$ correspond to thermally induced local 
nonaffine modes related to defect precursors in crystalline materials in both two and three 
dimensions. In disordered systems, large local $\chi_{i}$ is associated with shear transformation zones 
or soft spots, areas with a heightened tendency for irreversible particle 
rearrangements~\cite{Falk_Langer, DiDonna_Lubensky, Milkus_Zaccone}.

As a 2D granular medium with N particles undergoes jamming and unjamming with varying packing fraction 
$\phi$ at a finite strain rate $\dot{\varepsilon}_{\rm v}$, we track the cumulative nonaffine 
displacements for large, $\mathcal{X}_{\mathrm{L}}$ and small particles, $\mathcal{X}_{\mathrm{S}}$,
defined by $\mathcal{X}_{\mathrm{L}} = \sum_{i=1}^{N_{\mathrm{L}}} \chi^{\mathrm{L}}_{i}$ and 
$\mathcal{X}_{\mathrm{S}} = \sum_{i=1}^{N_{\mathrm{S}}} \chi^{\mathrm{S}}_{i}$, respectively.
$\chi^{\mathrm{L,S}}_{i}$ denotes the nonaffine displacement of each $i$th large or 
small particle in the system. Figs.~3 (e)-(f) in the paper illustrate the evolution of these 
variables as a function of $\hat{\phi}$ for monodisperse and bidisperse packings.


\bibliographystyle{unsrt}
\bibliography{RefSuplement}